\documentclass[twocolumn,showpacs]{revtex4}

\usepackage{graphicx}
\usepackage{dcolumn}
\usepackage{amsmath}
\usepackage{longtable}
\begin{document}

\title{Doping-Driven Collapse of the SDW Correlation Gap in SmFeAsO$_{1-x}$F$_{x}$}

\author{Scott~C.~Riggs$^1$, R.D.~McDonald$^2$, J.B.~Kemper$^1$, Z.~Stegen$^1$, G.S.~Boebinger$^1$, F.F.~Balakirev$^2$, Y.~Kohama$^2$, A.~Migliori$^2$, H.~Chen$^3$, R.H.~Liu$^3$ and X.H.~Chen$^3$}

\affiliation{$^1$National High Magnetic Field Laboratory, Florida State University, Tallahassee, FL, 32310, USA}
\affiliation{$^2$National High Magnetic Field Laboratory, Los Alamos National Laboratory,
MS-E536 Los Alamos, NM, 87545, USA}
\affiliation{$^3$Hefei National Laboratory for Physical Sciences at Microscale and Department of
Physics,University of Science and Technology of China, Hefei, Anhui 230026, P.R. China}

\begin{abstract}
We report the Hall resistivity, $\rho_{xy}$ of polycrystalline SmFeAsO$_{1-x}$F$_{x}$ for four different fluorine concentrations from the onset of superconductivity through the collapse of the structural phase transition. For the two more highly-doped samples, $\rho_{xy}$ is linear in magnetic field up to 50~T with only weak temperature dependence, reminiscent of a simple Fermi liquid. For the lightly-doped samples with $x<0.15$, we find a low temperature regime characterized $\rho_{xy}(H)$ being both non-linear in magnetic field and strongly temperature dependent even though the Hall angle is small. The onset temperature for this non-linear regime is in the vicinity of the structural phase (SPT)/spin density wave (SDW) transitions. The temperature dependence of the Hall resistivity is consistent with a thermal activation of carriers across an energy gap.  The evolution of the energy gap with doping is reported.

\end{abstract}


\date{\today}
\maketitle

An important key to unraveling the mystery of superconductivity in the ferropnictides will be understanding the phase diagram under different tuning parameters, including doping, pressure, and rare earth substitution.  For example, for five different rare earths (Re = La \cite{La1}, Nd \cite{Nd1}, Ce \cite{Ce1}, Pr \cite{Pr1}, and Sm \cite{Sm1}) the ReFeAsO$_{1-x}$F$_x$ (oxypnictide) system undergoes a structural phase transition (SPT) from tetragonal to orthorhombic symmetry.  Approximately 25K below the SPT the onset of long range magnetic order (LRO) is ascribed to spin density wave (SDW) formation \cite{SDW} . 

The transition temperatures for both the SPT and SDW decrease with increasing doping.  Above a critical doping, which is $x \approx 0.15$ for the Sm compound, there is no evidence of either the SPT or SDW transition.  Suppression of the energy scale of the SPT/SDW is observed not only with doping but also with increasing pressure \cite{Pressure1} or decreasing size of the rare earth ion \cite{RareEarth2}, \cite{RareEarth1}.  When doping is used to suppress the structural phase transition an insulator-to-metal crossover (IMC) is revealed at $x \approx 0.15$ once high magnetic fields are used to suppress superconductivity in SmFeAsO$_{1-x}$F$_x$ polycrystals \cite{Riggs1}.  The IMC has been confirmed in both CeFeAsO$_{1-x}$F$_x$ \cite{Singleton1} and BaFe$_2$As$_2$ \cite{Fu1} single crystals, and LaFeAsO$_{1-x}$F$_x$ polycrystals \cite{Yoshi1}, implying this phenomenon is general for the ferropnictide systems. Furthermore, the application of pressure to the undoped parent compound induces superconductivity yielding a phase diagram similar to that where chemical doping is the tuning parameter \cite{SPTPress1,SPTPress2}. Theoretical predictions argue that doping, rare earth substitution and pressure control the onset of pairing by suppressing long range magnetic order \cite{Mazin}.  To better understand how tuning the chemical potential suppresses long range order, we probe the doping dependent phase diagram of SmFeAsO$_{1-x}$F$_x$, and report the temperature and high field dependence of the Hall resistivity in a series of four samples with fluorine doping (F-doping) $0.05 \leq x ²\leq 0.20$.

For this investigation, the polycrystalline samples were synthesized using the conventional solid state reaction \cite{Sm1} and cut into rectangular prisms with a typical size of $1.5 \times 1 \times 0.1$~mm$^3$. The series of samples spans a large portion of the underdoped superconducting regime up to maximum $T_{\rm C}$ with  $x\approx 0.20$ \cite{Sm1} with transition temperatures measured at the midpoint of the superconducting transition, $T_{\rm C} \approx$ 2~K, 18~K, 40~K, and 46~K for $x=0.05$, $x=0.12$, $x=0.15$ and $x=0.20$ respectively. These are the same series of samples in which the insulator-to-metal crossover was reported \cite{Riggs1}.

The Hall resistivity was measured using a standard four-terminal digital ac lock-in technique with the magnetic field applied normal to the large face of the sample using pulsed fields up to 50~T and continuous fields up to 35~T at the National High Magnetic Field Laboratory. For pulsed field measurements, at each temperature the resistivity transverse to the magnetic field and applied current was measured during two magnetic field pulses of opposite polarity to subtract any contamination of the longitudinal resistivity from the Hall resistivity.  

\begin{figure}
\centering
\includegraphics[width=8.2cm]{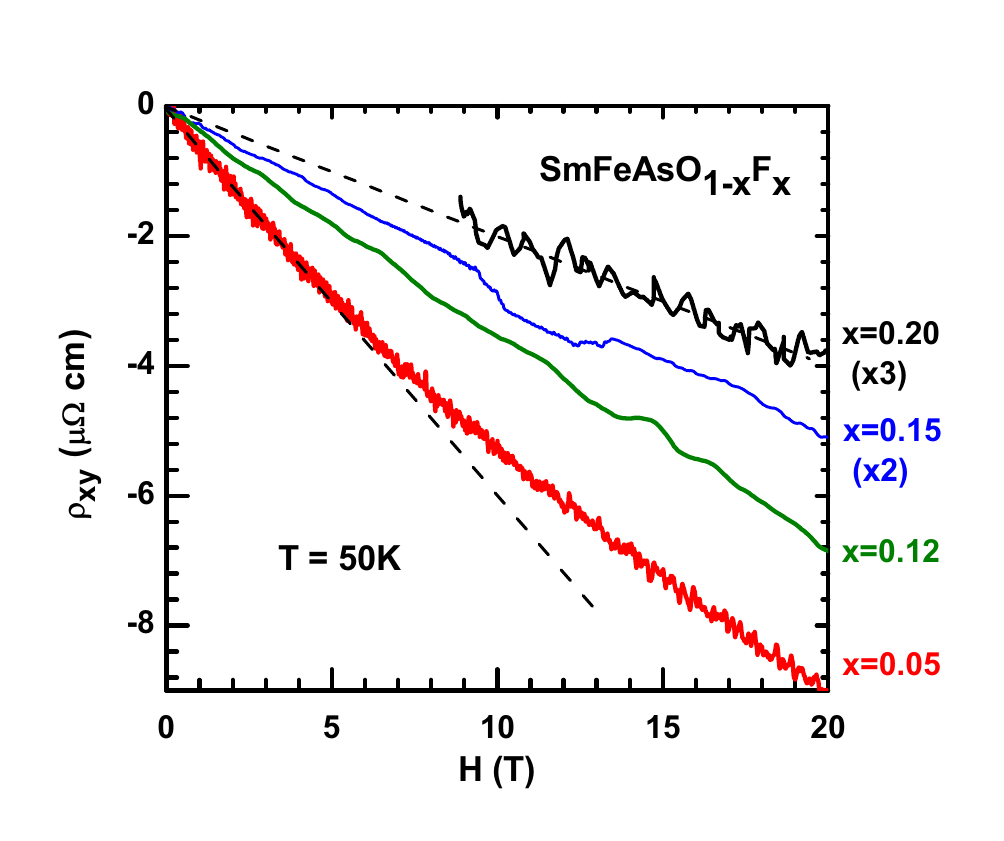}
\caption{(color online)  The transverse magneto-resistance, $\rho_{xy}(H)$ for compositions with $x = 0.20, 0.15, 0.12$ and $0.05$ (top to bottom) measured at a temperature of 50~K. The non-linearity of  $\rho_{xy}$ for the lowest  doping is clearly observed.  The dashed lines are linear fits to at low fields.}
\label{Fig1}
\end{figure}

\begin{figure}
\centering
\includegraphics[width=8.2cm]{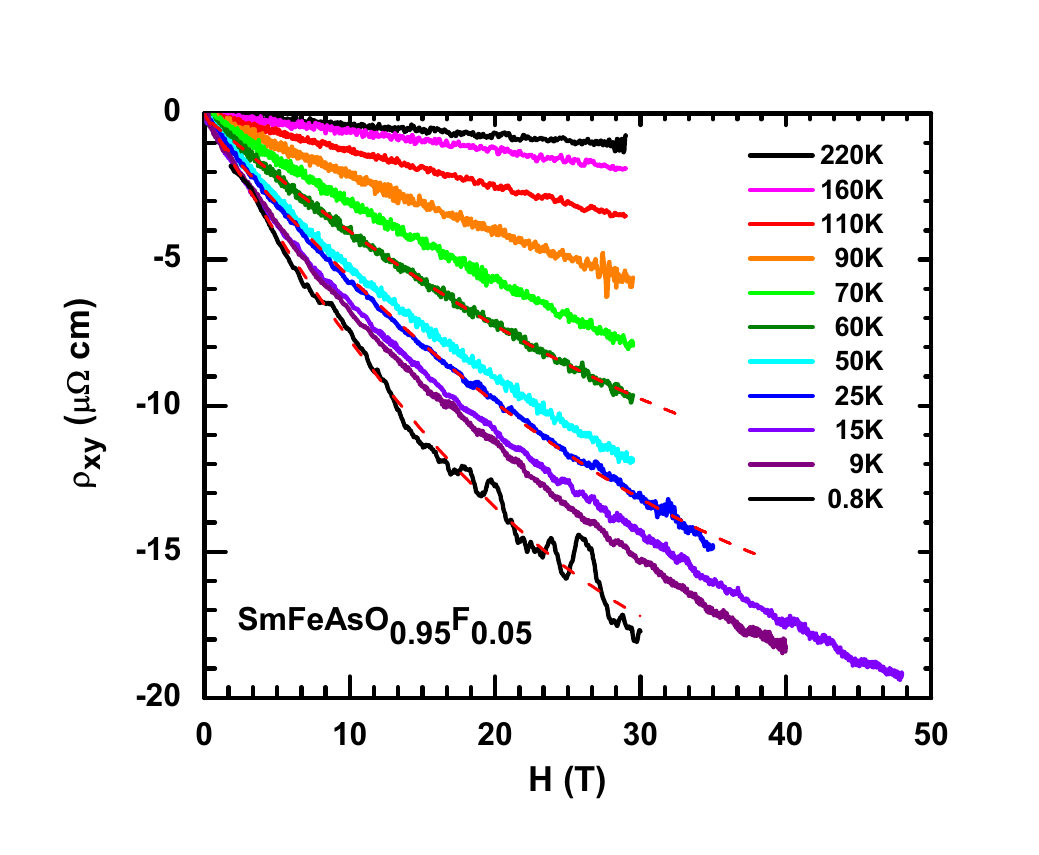}
\includegraphics[width=8.2cm]{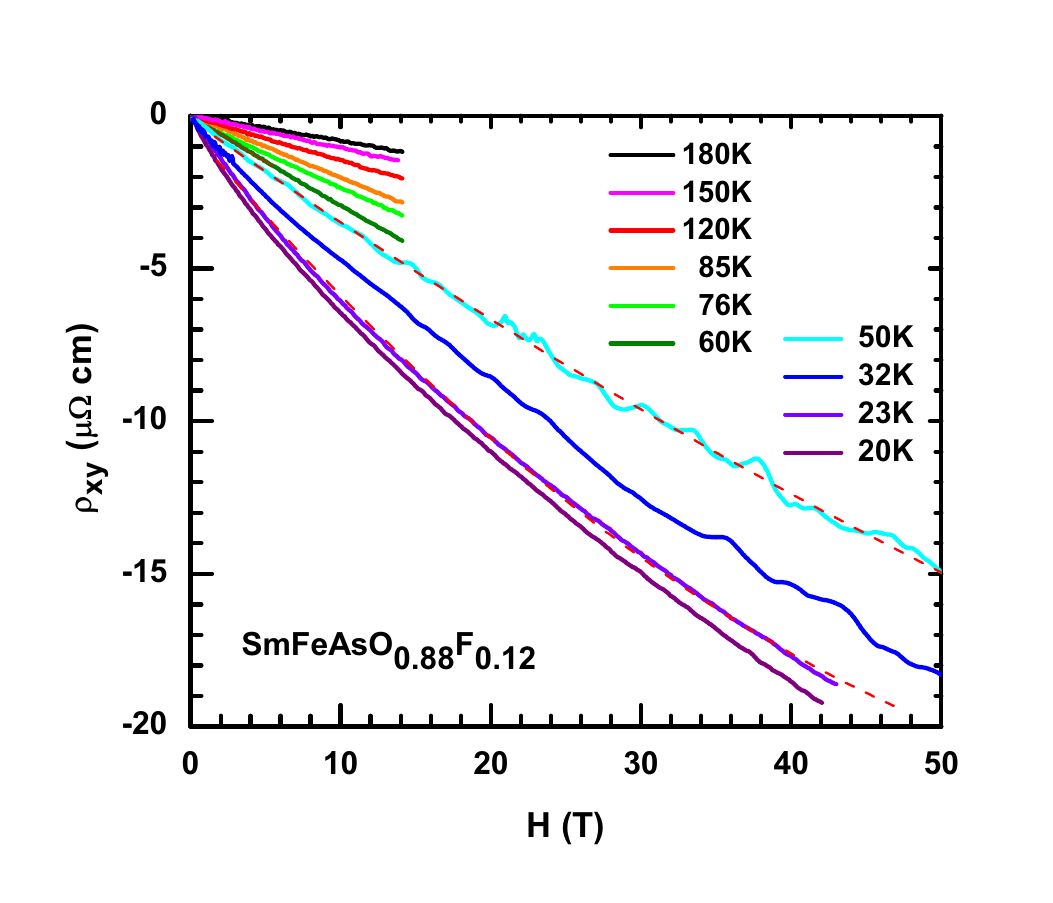}
\includegraphics[width=8.2cm]{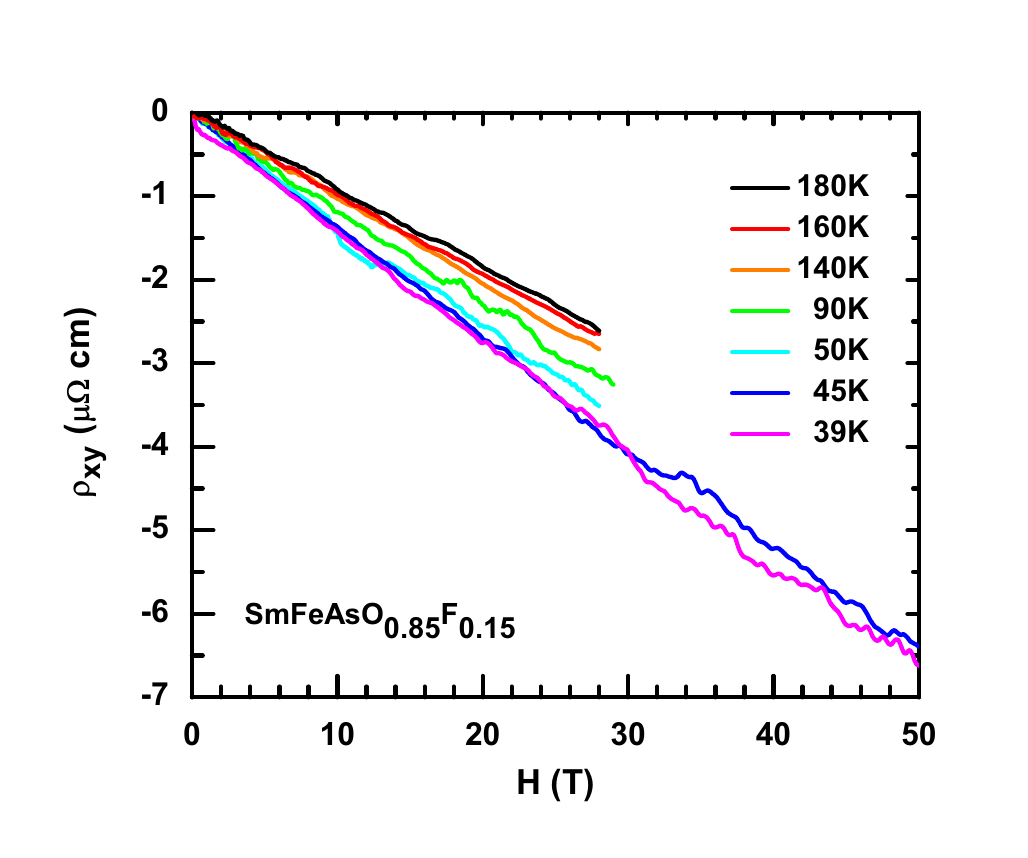}
\caption{(color online) The evolution of $\rho_{xy}(H)$ with temperature for $x = 0.05$ (top), $x = 0.12$ (middle) and $x = 0.15$ (bottom). Note that both the non-linear $\rho_{xy}(H)$ and temperature dependence weaken with increasing doping.  Dashed lines are fits to Equation~\ref{pheno1}.}
\label{Fig2}
\end{figure}

Figure~\ref{Fig1} contains $\rho_{xy}(H)$ traces for all four samples at a fixed temperature of 50~K up to 20~T. 
The small values of the Hall angle \cite{SDW} reported over the entire temperature and doping range of this investigation, are consistent with being in the the low field limit , $\omega_{\rm C}\tau << 1$ \footnote{where  $\omega_{\rm C}$ is the cyclotron frequency and $\tau$ the scattering rate},  i.e. the regime of a linear dependence of the Hall voltage in magnetic field; $\rho_{xy}(H) = R_{\rm Hall}H$ with $R_{\rm Hall}^{-1}$ yielding the total number of carriers that contribute to transport. At a temperature of 50~K, linear $\rho_{xy}(H)$ is observed for all doping levels except $x=0.05$, i.e. the sample with the lowest doping, see Figure~\ref{Fig1}. In order to probe a wider regime, we measure Hall resistivity up to 50~T over a broad temperature range.   For the most highly doped sample, $x=0.20$, $\rho_{xy}(H)$ is linear in field for all temperatures measured, consistent with our previous report for another doping above the observed disappearance of the SPT, $x=0.18$ \cite{Riggs2}.  Figure~\ref{Fig2} depicts the magnetic field dependence of the Hall resistivity  $\rho_{xy}(H)$ for the other three dopings. The Hall resistivity is negative for all samples at all temperatures measured, evidencing the dominance of electron-like charge carriers, as is expected for F-doping the parent compound, a compensated semi-metal \cite{Nd1}.   
  
The most striking features of the data in Figures~\ref{Fig1} and \ref{Fig2} are: (a) that the temperature dependence of $\rho_{xy}$ increases as doping is decreased and (b) that all non-linear behavior in $\rho_{xy}$ is observed at temperatures below the reported SPT/SDW transition, and (c) the non-linearity grows with decreasing temperature.   The effect of temperature on $\rho_{xy}$ is the most dramatic at $x=0.05$, the least doped sample studied.  The temperature dependence of $\rho_{xy}$ decreases with doping until almost no temperature dependence is observed in $\rho_{xy}$ for $x = 0.20$.   

\begin{figure}
\centering
\includegraphics[width=8.2cm]{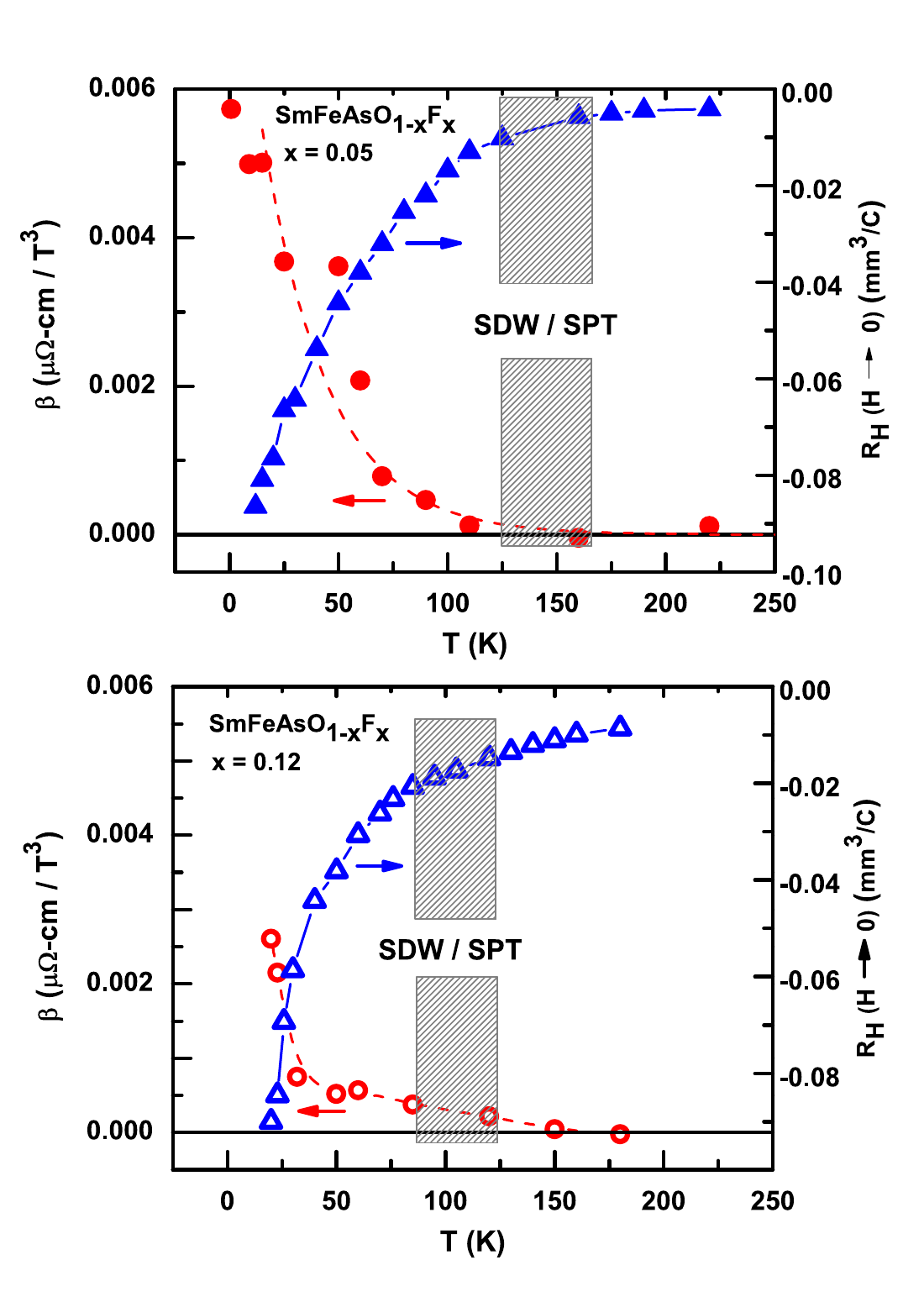}
\caption{(color online) Temperature dependence of $R_{\rm Hall}$ in the low-field limit (triangles) and
$\beta$, the non-linear term of Equation~\ref{pheno1} (circles) for x = 0.05 (top) and x=0.12 (bottom).
There is a strong temperature dependence of both quantities below the
SDW/SPT transitions.  In particular, at high temperatures, $\beta = 0$, implying linear $\rho_{xy}( H )$ until the onset of the
SDW/SPT.  At lower temperatures, the non-linearity of $\rho_{xy}(H)$ grows dramatically with decreasing 
temperature. Dashed lines are guides for the eye.}
\label{Fig3}
\end{figure}

Non-linearity can arise in semi-metals because of complexity in transport of a multiband system \cite{Kittel}.  We characterize the extent of the nonlinear regime in the temperature-doping phase diagram by fitting the Hall data in Figures~\ref{Fig1}a and \ref{Fig1}b to a phenomenological formula (dashed lines): 
\begin{equation}
\label{pheno1}
\rho_{xy}(H) = R_{\rm Hall}H - \beta H^3
\end{equation} 
We plot both the linear and cubic coefficients for $x=0.05$ and $0.12$ in Figure~\ref{Fig3}. Note that both $R_{\rm Hall}(T)$ and $\beta(T)$ exhibit weak temperature dependence at high temperatures where  $\beta \approx 0$  {\it i.e.} the regime of conventional Hall behavior as seen at all measured temperatures in the $x =0.15$, $0.18$, and $0.20$ materials. However, for the two lowest dopings, shown in Figure~\ref{Fig3}, both $|R_{\rm Hall}(T)|$ and $|\beta(T)|$ grow dramatically as the temperature is decreased below the SPT/SDW (shaded region).  This is the signature of the Ônon-linearÕ regime. 
The non-linearity in magnetic field is consistent with a SDW transition removing sections of Fermi surface, and also decreasing the scattering rate (increasing $\tau$) once long range order has frozen out the fluctuations of the nesting vector. 

\begin{figure}
\centering
\includegraphics[width=8.2cm]{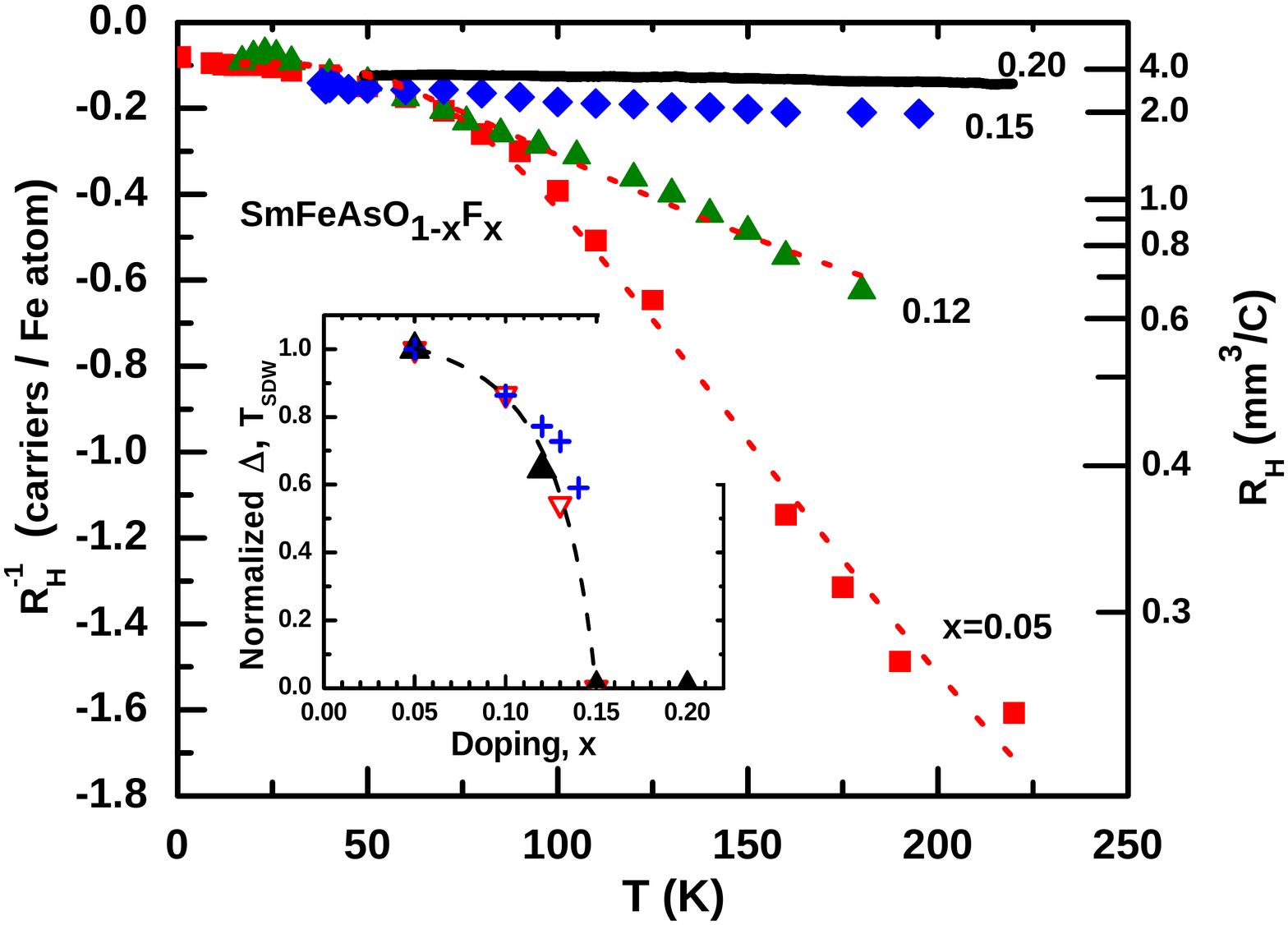}
\caption{(color online) Temperature dependence of the Hall number (right) and inverse Hall number, normalized
to carriers per Fe-atom (left). The dashed lines are fits to Equation~\ref{Sigma1}. The inset shows the doping evolution of the energy gap, normalized to the value of $\Delta$ at x=0.05 for this study (black up triangles) and Liu,~{\it et al}~\cite{SDW} (red down triangles).   For comparison, the doping dependence of the spin density wave transition (blue crosses), $T_{\rm SDW}$, is also provided, normalized to $T_{\rm SDW} (x=0.05) = 114$~K  \cite{SDW}.}
\label{Fig4}
\end{figure}

We now examine the temperature dependence of the data in Figures~\ref{Fig1} and \ref{Fig2} separately from the non-linearity in magnetic field, i.e. we consider only the low field data, where all data are linear in $H$.  Figure~\ref{Fig4} shows the inverse Hall coefficient for all four doping levels normalized as carriers per Fe atom and $cm^{3}/C$. At low temperatures, $|R_{\rm Hall}^{-1}|$  is comparable to the level of F-doping, increasing exponentially with temperature (dotted lines) for samples in the Ônon-linearÕ regime. The exponential dependence suggests thermal excitation of carriers. In the low-field limit, the Hall coefficient in a two-band model \cite{Kittel}, reduces to: 
\begin{equation}
\label{Rhall1}
R_{\rm Hall} (H\rightarrow0) = (-1)\biggl[\frac{\sigma_0\mu_0 \pm \sigma_1\mu_1}{(\sigma_0 + \sigma_1)^2}\biggr]
\end{equation} 
where both $\sigma_i$ and $\mu_i$ are positive and Ô0Õ denotes the (electron-like) behavior of the semi-metal, and Ô1Õ represents additional thermally activated carriers respectively, the -/+ denotes whether the activated carriers are electrons or holes respectively.  As such, we propose that
\begin{equation}
\label{Sigma1}
\sigma_1(T) = \frac{e^2\tau}{m^{\star}_1}n_{1}(T) = \frac{e^2\tau}{m^{\star}_1}n_{1}^{0}\exp \biggl\{\frac{-\Delta}{k_{\rm B}T}\biggr\}, 
\end{equation} 
where $\sigma_1\rightarrow0$ as $T\rightarrow0$.
At low temperatures and to first order in $\sigma_1\slash\sigma_0$, the inverse Hall coefficient
\begin{eqnarray}
\label{Rhall2}
R_{\rm Hall}^{-1} (H\rightarrow0) = (-1)\biggl(\frac{\sigma_0}{\mu_0}\biggr)\biggl[1+\frac{\sigma_1}{\sigma_0} \mp \frac{\mu_1\sigma_1}{\mu_0\sigma_0}\biggr] 
\nonumber \\
 = \biggl(-|ne|\biggr)\biggl[1+(1+2\alpha \mp \alpha^2) \frac{n_1(T)}{n_0}\biggr] 
\end{eqnarray} 

where $\alpha = \mu_1\slash\mu_0$, thus $\sigma_1\slash\sigma_0 = \alpha n_1(T)\slash n_0$.  
Fitting  (dashed lines in Figure~\ref{Fig4}) yields a measure of the energy gap $\Delta$.  We determine gap values of $\Delta = 24.9$~meV for $x=0.05$ and $\Delta = 16.1$~meV for $x=0.12$,which is in excellent agreement with the $\approx13$~meV reported from photoemission on SmFeAsO$_{0.88}$F$_{0.12}$ and SmFeAsO$_{0.85}$F$_{0.15}$ \cite{SmARPES1,SmARPES2}.  

The energy scale of this gap is too small to reconcile with the predictions for the unreconstructed (bare) band structure\cite{BS}, but is comparable to the SDW ordering temperature $(\Delta \approx k_{\rm B}T_{\rm SDW})$ indicating we are likely probing the evolution of a gap arising from electron correlations as a function of doping. The long range SDW order most likely arises because of nesting the zone center hole sections of Fermi surface with the zone corner electron pockets \cite{Amalia1}. Within this model one can naturally understand how the Fermi surface reconstruction is sensitive to doping and pressure:  doping suppresses nesting by unbalancing the electron and hole pockets, while pressure (chemical - rare earth substitution or hydrostatic) increases the electronic band width and interlayer warping, which in a multiband system suppresses nesting.

The inset to Figure~\ref{Fig4} plots the activation energy $\Delta$ extracted from the temperature dependence of $\rho_{xy}$ for the compositions of this investigation and those reported by Liu~{\it et al} \cite{SDW}.
Both data sets clearly demonstrate a suppression of the activation energy with doping that collapses onto a single curve (dashed line) when normalized by the gap energy for the common doping ($x = 0.05$, at which $\Delta \approx 24.9$~meV for this investigation and $\Delta \approx 44.6$~meV from the data in Lui, et al). 

In the paradigm SDW system, Chromium and its alloys \cite{CrStrain} it has been shown that the square of the amplitude of the density wave is proportional to the strain amplitude. Evidence for this sensitivity to strain is also mounting for the ferropnictide family in both the strong hydrostatic pressure dependence of the phase diagram \cite{Pressure1} and the discrepancy between the Fermi surface topology as derived from bulk measurements \cite{Amalia1} and ARPES surface studies \cite{ZXShen}.  In particular, mechanical thinning of the samples is required for pulsed magnetic field measurements in order to avoid sample heating and increase the Hall signal, which is likely to cause a differing degree of strain than is present in the `as grown' samples.  Given our interpretation of the gap as originating from the SDW ordering, it is not surprising that such normalization is necessary to account for the extreme sensitivity of the gap to strain caused by sample preparation and handling.

From the inset of Figure~\ref{Fig4} it is evident that both $T_{\rm SDW}$ and $\Delta$ are suppressed with doping. It is also evident that the gap is being suppressed more rapidly than the transition temperature, such that an estimate of the coupling strength as the ratio of energy gap to the transition temperature\cite{PPL,BCS}, yields a value of  $\approx 5$ (typical of the strong coupling found in density wave systems \cite{PPL}) for the x=0.05 sample of Lui et al and a value of $\approx 4$ for $x = 0.13$. 

In conclusion, we have defined a small-Hall-angle regime in SmFeAsO$_{1-x}$F$_x$ that is characterized by unusual behavior of $\rho_{xy}(H)$: nonlinear in magnetic field and exponential in temperature. This regime exists at low fluorine doping, $x < 0.15$, and at temperatures below the structural and SDW phase transitions. At either higher temperatures or doping, the Hall resistivity behaves like a conventional metal. Finally, the demarcation at $x\approx0.15$ between the nonlinear regime and conventional metal behavior is the same doping where there is both an end to the SPT/SDW ordering and an insulator-to-metal crossover in the normal state.  

Part of this work was supported by NSF Cooperative Agreement No. DMR-0654118, by the State of Florida, and
by the DOE.  Scott~C.~Riggs would like to acknowledge the ICAM travel fellowship for financial support.  We would like to thank Lev~Gor'kov, Z.~Tesanovic, I.~Mazin, J.~Analytis, and I.R.~Fisher for stimulating discussions.

\newpage

\begin{thebibliography}{Bibliography}
 

\bibitem{La1}
C.~de la Cruz {\it et al}, Nature 453,(2008). {\it Magnetic order close to superconductivity in the iron-base layered  LaFeAsO$_{1?x}$F$_x$  systems.}

\bibitem{Nd1}
Michela Fratini et al, {\it The effect of internal pressure on the tetragonal to monoclinic structural phase transition in ReOFeAs: the case of NdOFeAs} Supercond. Sci. Technol. 21 092002 (2008)

\bibitem{Ce1}
Jun Zhao, Q. Huang, Clarina de la Cruz, Shiliang Li, J. W. Lynn, Y. Chen, M. A. Green, G. F. Chen, G. Li, Z. Li, J. L. Luo, N. L. Wang, Pengcheng Dai, {\it Structural and magnetic phase diagram of CeFeAsO$_{1-x}$F$_x$ and its relationship to high-temperature superconductivity}, Cond-mat, arXiv: 0806.2528

\bibitem{Pr1}
J.~Zhao, Q.~Huang, C.~de la Cruz, J.W.~Lynn, M.D.~Lumsden, Z.A.~Ren, J.~Yang, X.~Shen, X.~Dong, Z.~Zhao, and P.~Dai, Cond-mat, arXiv: 0807.4872

\bibitem{Sm1}
X.H. Chen, T. Wu, G. Wu, R.H. Liu, H. Chen and D.F. Fang, Nature 453,
761(2008). {\it Superconductivity at 43 K in Samarium-arsenide Oxides: SmFeAsO$_{1?x}$F$_x$}

\bibitem{SDW}
R.H.~Liu, F.~Wu, T.~Wu, D.F.~Fang, H.~Chen, S.Y.~Li, K.~Liu, Y.L.~Xie, X.F.~Wang, R.L.~Yang, L.~Ding, C.~He, D.L.~Fang, and X.H.~Chen, Cond-mat, arXiv: 0804.2105

\bibitem{Pressure1}
Kazumi Igawa et al, Cond-mat, arXiv: 0809.1239

\bibitem{RareEarth2}
Karolina Kasperkiewicz, Jan-Willem G. Bos, Andrew N. Fitch, Kosmas Prassides, Serena Margadonna, Cond-mat, arXiv:  0809.1755

\bibitem{RareEarth1}
Zhi-An Ren et al, {\it Superconductivity and phase diagram in iron-based arsenic-oxides ReFeAsO$_{1-\delta}
$ (Re-rare-earth metal) without Flourine doping}, EPL 83 17007 (2008).

\bibitem{Riggs1}
S.C.~Riggs, J.B.~Kemper, Y.~Jo, Z.~Stegen, L.~Balicas, G.S.~Boebinger, F.F.~Balakirev, A.~Migliori, H.~Chen, R.H.~Liu, X.H.~Chen, Cond-mat, arXiv: 0806.4011

\bibitem{Singleton1}
J. Singleton, {\it Private communication} (2008)

\bibitem{Fu1}
 H. Q. Yuan, J. Singleton, F. F. Balakirev, G. F. Chen, J. L. Luo, N. L. Wang, Cond-mat, arXiv: 0807.3137

\bibitem{Yoshi1}
Y. Kohama, Y. Kamihara, S. A. Baily, L. Civale, Scott C. Riggs, F. F. Balakirev, T. Atake, M.
Jaime, M. Hirano, and H. Hosono, Cond-mat, arXiv: 0809.1133


\bibitem{SPTPress1}
Patricia L. Alireza, Jack Gillett, Y. T. Chris Ko, Suchitra E. Sebastian, Gilbert G. Lonzarich, Cond-mat, arXiv: 0807.1896

\bibitem{SPTPress2}
M. S. Torikachvili, S. L. Bud'ko, N. Ni, P. C. Canfield, Cond-mat, arXiv: 0809.1080

\bibitem{Mazin}
I.I. Mazin {\em et al.},
arXiv: 0807.3737

\bibitem{Riggs2}
J Jaroszynski, Scott C. Riggs, F. Hunte, A. Gurevich, D. C. Larbalestier, G. S. Boebinger, F. F. Balakirev, Albert Migliori, Z. A. Ren, W. Lu, J. Yang, X. L. Shen, X. L. Dong, Z. X. Zhao, R. Jin, A. S. Sefat, M. A. McGuire, B. C. Sales, D. K. Christen, and D. Mandrus, Phys. Rev. B 78, 064511 (2008)

\bibitem{Kittel}
C.Kittle, Introduction to Solid State Physics, Weiley and Sons ( 7th edition) 1996.

\bibitem{SmARPES1}
Haiyun Liu et al,  {\it Supercondting Gap and Pseudogap in SmO$_{1-x}$F$_x$FeAs Layered Superconductor from Photoemission Spectroscopy}, Cond-mat, arXiv:0805.382

\bibitem{SmARPES2}
Xiaowen Jia et al, {\it Common Features in Electronic Structure of the Fe-Based Layered Superconductors from Photoemission Spectroscopy}, Cond-mat, arXiv:  0806.0291

\bibitem{BS}
D.~J.~Singh and M.-H.~Du Phys Rev. Lett. 100, 237003 (2008) 

\bibitem{Amalia1}
A.I. Coldea, J.D. Fletcher, A. Carrington, J.G. Analytis, A.F. Bangura, J.-H. Chu, A.S. Erickson, I.R. Fisher, N.E. Hussey, R.D. McDonald, { \it Fermi surface of a ferrooxypnictide superconductor determined by quantum
oscillations}, Cond-mat, arXiv: 0807.4890 

\bibitem{CrStrain}
S.~Iida, M.~Kohno, Y.~Tsunoda and N.~Kuntiomi, J.~Phys. Soc. Japan {\bf 44} 1747 (1978). 

\bibitem{ZXShen}
D.H.~Lu {\em et al.}, arXiv:0807.2009.

\bibitem{PPL} 
R.D.~McDonald, N.~Harrison, L.~Balicas, K.H.~Kim, J. Singleton and X.~Chi, Phys. Rev. Lett {\bf 93}, 076405, (2004) 

\bibitem{BCS}
J.~Bardeen, L.N.~Cooper and J.R.~Schrieffer, Phys. Rev. {\bf 108}, 1175 (1957).
 
\end{thebibliography}
\end{document}